\documentclass[a4paper]{jpconf}
\usepackage{caption}
\usepackage{subcaption}
\usepackage{graphicx}
\usepackage{epsfig}
\usepackage{epstopdf}
\usepackage{array}
\usepackage{siunitx}
\usepackage[usenames, dvipsnames]{color}
\usepackage{color}
\usepackage{longtable}
\usepackage[utf8]{luainputenc}
\usepackage{amssymb}

\usepackage{xspace}
\usepackage{multicol}

\usepackage{makecell}
\usepackage{cite}

\newcommand{\Smilei}{{\sc Smilei}\xspace}
\usepackage{enumitem}

\usepackage{svg}

\begin{document}
\title{Consequences of laser transverse imperfections on laser wakefield acceleration at the Apollon facility}

\author{Imene Zemzemi$^1$, Arnaud Beck$^1$, Arnd Specka$^1$}

\address{$^1$ Laboratoire Leprince-Ringuet – École polytechnique, CNRS-IN2P3, Palaiseau 91128, France}
\ead{beck@llr.in2p3.fr}

\begin{abstract}
With the currently available laser powers, it is possible to reach the blowout regime in the Laser WakeField Acceleration (LWFA) where the electrons are completely expelled off-axis behind the laser pulse. This regime is particularly interesting thanks to its linear focusing forces and to its accelerating forces that are independent of the transverse coordinates. In fact, these features ensure a quite stable propagation of electron bunches with low phase-space volume. In this context, the Apollon laser is designed to reach an exceptional multi-petawatt laser peak power, thus aiming at achieving unprecedented accelerating gradients and bringing a scientific breakthrough in the field of LWFA. 

Since the quality of the self-injected electron bunches is very sensitive to the condition of the laser, it is very important to take into account 
realistic laser features when performing LWFA simulations. In this paper, we aim at understanding the implications of laser imperfections on the electrons produced with the self-injection scheme in the bubble regime. For this purpose, we carry on a numerical study of LWFA where we include experimentally measured laser profiles from the Apollon facility in full three dimensional Particle-In-Cell simulations.

\end{abstract}

\section{Introduction}\label{intro}

Laser wakefield acceleration (LWFA) has been demonstrated as an established technique for accelerating electrons efficiently  via the
interaction of a femtosecond-scale laser pulse with a plasma.  The continuous  progress in increasing the laser power made it possible to  exceed the critical power for relativistic self-focusing ($P/P_c >1$) \cite{Sun87}, hence  the laser pulses can be  self-focused over several Rayleigh lengths. The region where the laser is self-focused triggers  a highly nonlinear plasma wave in which plasma electrons are trapped and accelerated to relativistic energies in millimeter distances. Nowadays, the highest acceleration gradients are reached in the bubble regime allowing the generation of multi-GeV electrons. Owing to its simplicity,  self-injection is definitely the least experimentally demanding \cite{Benedetti2013}, thus the most commonly used mechanism to produce multi-GeV electrons \cite{ Malka2013proc}. 

However,  the self-injection technique offers little control over the injection process  because it  is very sensitive to both the conditions of the plasma and laser.  The injection process is a key factor to determine the final bunch properties.  This emphasizes the importance of establishing  a  more stable and controllable self-injection in order to generate  laser-driven electron bunches that meet the requirements in terms of high charge, low emittance, and low-energy spread simultaneously. 

The improvement of this injection scheme relies intrinsically on understanding the sensitivity of the accelerator's performance to deviations from the ideal physics. 
Specifically, it is known that the laser pulse deviation from the ideal Gaussian assumption is critical to the bunch quality. Therefore, shot-to-shot fluctuations of the laser intensity may impact drastically the reproducibility of the electrons bunch in terms of beam charge, energy, and emittance. 
In fact, tightly focused femtosecond laser pulses used  to accelerate the electrons from the plasma have  a complicated structure with asymmetries  in the laser's focal spot, aberrations in the wavefront, and variations in the temporal profile of the pulse. These imperfections affecting the laser pulse propagation \cite{Kaluza2010, Glinec2008}, also affect the self-injection processes and may lead to poor bunch quality \cite{Ferri2016, Vieira2012}. Nevertheless, it has been demonstrated that laser profiles exhibiting  higher-order  Laguerre-Gauss modes may lower the  self-trapping thresholds when they are properly tailored \cite{Michel2006} and that  laser profile imperfections can promote the production of  betatron oscillations  \cite{Glinec2008, Mangles2009,Ferri2016}.

Kinetic simulations provide an ideal venue to investigate and help understand the response of the LWFA to  laser imperfections as it allows deliberate control of the shape of the introduced laser pulse and its associated phase. 

So far, most of the theoretical and numerical studies dealing with realistic laser pulses used either higher-order  Laguerre–Gaussian transverse profiles \cite{Vieira2012, Genoud2013} or  theoretical fits  of the experimental measure. For example, a combination of  two Gaussian pulses with the same duration was used  in \cite{Nakanii2016}  to mimic the halo effect by shifting the  center of  the second lower-energy pulse from the axis of the first pulse in the transverse direction. In \cite{Maslarova2019}, the laser beam has been chosen  to fit a profile that reaches the super-Gaussian  in the focus, and in \cite{Kirchen2021} the measured radial laser intensity evolution was approximated by a flattened Gaussian beam \cite{Gori1994}. We also find in
\cite{Cummings2011} that the presence of comatic aberration is modeled by modifying a Gaussian pulse and adding some terms from  the expansion of  Bessel functions  and Zernike Polynomials in the diffraction integral.

In this paper, the impact of realistic laser imperfections 
 on the electron bunch formation by the self-injection technique in the bubble regime, is studied  numerically via fully relativistic 3D particle-in-cell (PIC) simulations. 
In order to enlighten the origin of some trends that may be found in the future experiments of LWFA with the Apollon laser, the experimental  wavefront measured in the Apollon laser as well as its intensity profile are directly included in the simulations presented in this paper. Then, their influences are studied by comparing the results to that of a theoretical fit of the intensity profile with a super-Gaussian along with a flat wavefront.

In the following, first the experimental data acquisition and the numerical settings are described. Then, the simulation results of the laser propagation and focusing as well as the electron injection and acceleration are compared. Finally, the influence of the realistic parameters on the simulation is discussed. 

\section{Methods}\label{method}

\subsection{Experimental setup and data acquisition}\label{Exp}

\begin{figure}
\begin{center}
     \includegraphics[width=0.8\textwidth]{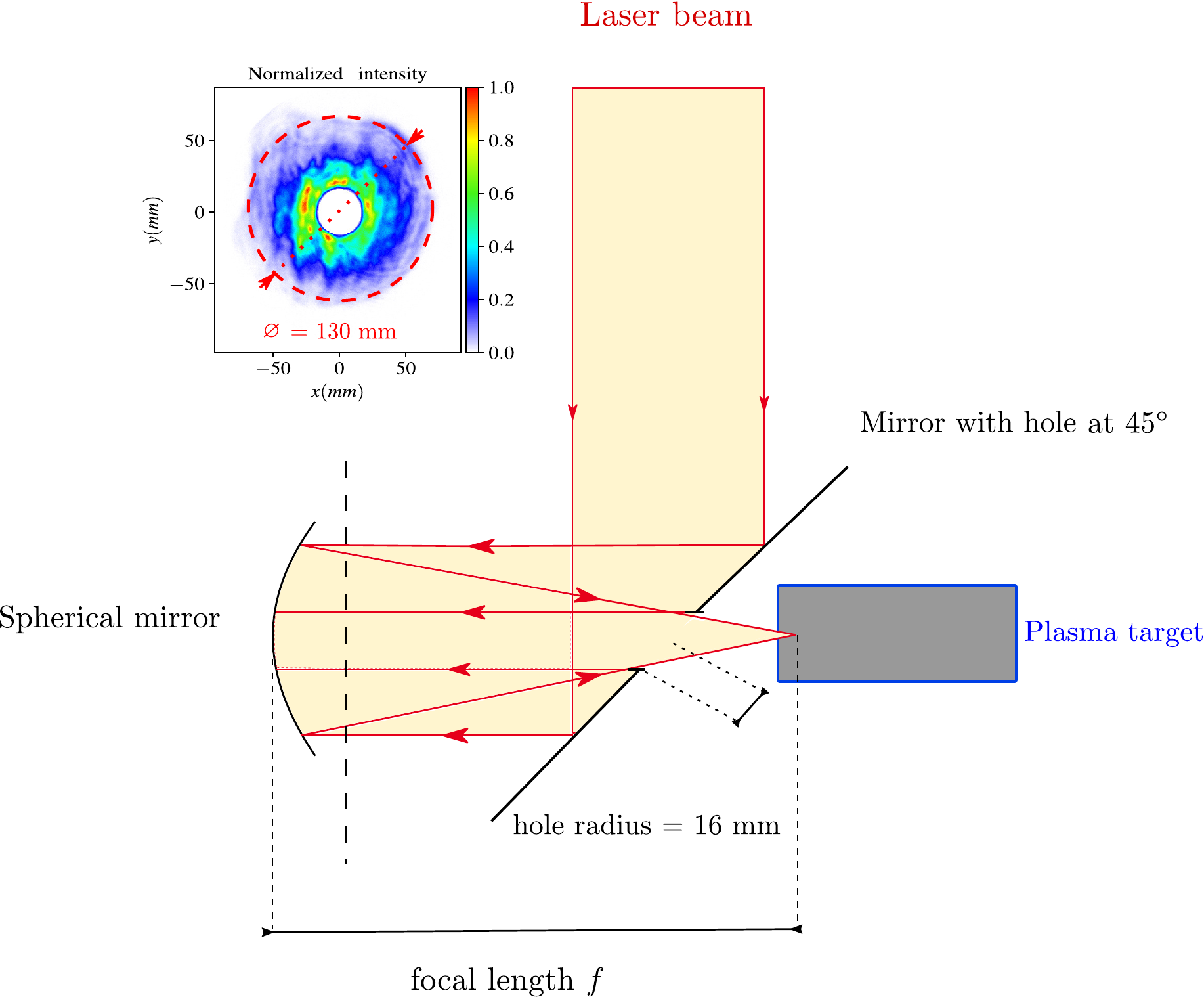}
\end{center}
  \caption{Schematic representation of the experimental set-up for the laser beam propagation and focusing in Apollon installation: a laser beam of a $1/e^2$ diameter $\simeq 130~ mm$ is sent and reflected  by a mirror at $45^{\circ}$ presenting a hole of a radius $ =16~ mm$ where a small fraction from the center of the laser is lost (estimated to be around $\sim 12 \%$). The beam is focused by a spherical mirror of a focal length $f$. The top left panel shows the intensity measured in the near field i.e. before the reflection on the spherical mirror.
 }
 \label{fig:experimental-setup}
\end{figure}

The experimental data measurement was carried out using the  Apollon laser facility in Saclay, France.  The laser system delivers a  linearly polarized pulse of $0.8~ \mu m$ wavelength  with a $25$ fs FWHM duration and total  energy of 15 J corresponding to 600 TW peak power. The measurements were obtained in the near field after the reflection on the mirror with the hole represented in figure \ref{fig:experimental-setup}.

 From shot to shot, the intensity profile variation in the near field is very small while the wavefront does fluctuate. Therefore, the wavefront measurement is averaged over  $10-100$ shots  to have a representative case of the Apollon beam systematic defaults.  Note that spatio-temporal distortions i.e  spatial dependencies of the temporal properties are not taken into consideration and that the wavefront is  also averaged over its spectrum.

The raw data  from the camera in the near field  is interpolated into the computational grid using a quadratic interpolation function taking into account the number of pixels of the camera and the pixel size.
Using Fresnel diffraction, the laser profile in the near field is then focused and  propagated up to the beginning of the plasma target where the PIC simulation starts at $z_{\rm start}= f -5 \times 10^{-4}~ m$. In the previous expression, $f$ denotes the focal length of the spherical mirror positioned at $z=0$.  Apollon facility offers the possibility of using two focal lengths of $3$ or $9 ~ m$. This  study is carried with $f = 3~m$.
Given the information about the laser intensity profile and phase,
Fresnel propagation allows an accurate calculation of the laser field at any point in the far field i.e. close enough to the focal plane.
In this case, it results in a tightly focused laser spot size  $w_0 \sim 15.5  \approx \mu m$ which corresponds to a full width half maximum in the intensity of $\rm FWHM_I \approx 25.8 ~ \mu m $. 

\subsection{Methodology}

For this study, we chose 3 scenarios for our simulations in order to understand the impact of realistic laser features on LWFA.
In the first scenario, the experimental measurement of the intensity laser profile is fitted with a super-Gaussian of order 4 and a flat wavefront is assumed. In the second case, the same super-Gaussian fit is used but with the measured wavefront. Finally, both the experimentally measured  intensity profile and wavefront are used. 

\begin{figure}[!h]
\centering
\includegraphics[width=\linewidth]{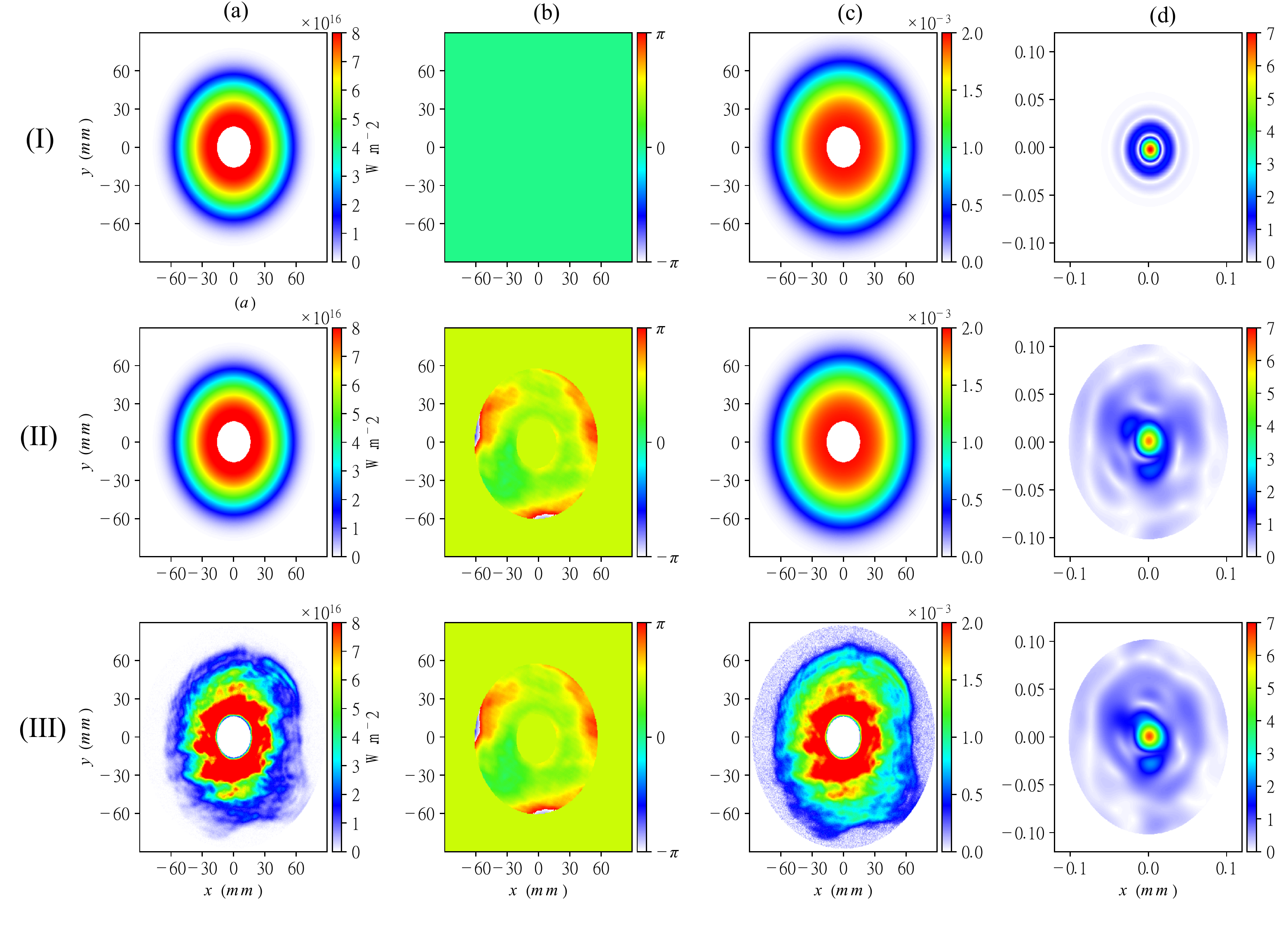}
\caption{Columns (a) and (b) present respectively the intensity  and the wave front transverse profiles in the near field. Column (c) shows the corresponding amplitude of the normalized vector potential $|a|$ in the near field. Column (d) shows the normalized vector potential $|a|$ in the far field, which is the results of Fresnel propagation of the field in column (c) up to $z=f-5 \times 10^{-4}~ m$ where $f=3$ m. The top line (I) corresponds to the case of a super-Gaussian fit $\mathrm{SG}_4$ for the intensity profile with a null wave front, the middle line (II) corresponds to the case of super-Gaussian fit $\mathrm{SG}_4$  of the intensity profile and the measured wavefront, and the bottom line (III) corresponds to the case of the  experimentally measured intensity profile and wavefront. }
\label{Near-far-profiles}
\end{figure}

Figure  \ref{Near-far-profiles} presents, for the 3 different cases, the intensity (column (a))  and the wave front (column (b)) colormaps and their corresponding normalized vector potential amplitude $|a|$ in the near field (column (c)). Finally, the same quantity is shown in the far field (column (d)) near the focal plan at $z=f-5 \times 10^{-4}~ m$ obtained numerically after a Fresnel propagation using a focal length of $f$=3 m. From this figure, we can see that the wave front is the key factor responsible on the main shape of the laser in the far field. 

 In the following, the three cases are referred to as: $\mathrm{SG}_{4}+\phi_{\mathrm{flat}}$ for the super-Gaussian fit of order 4 assuming a perfectly flat wavefront (null phase), $\mathrm{SG}_{4}+ \phi_{\mathrm{measured}}$ for the super-Gaussian fit of order 4  with the measured wavefront and $\mathrm{I}+ \phi_{\mathrm{measured}}$ for the experimentally measured intensity and wave front.

 In all three configurations, the  laser beams  have the same duration and  contain the same total energy which corresponds to the remaining energy after the reflection on the mirror with the hole where a fraction of the energy is lost $(\sim 12 \%)$.
 In all the three cases, the waists obtained after propagation are very similar. This tells that the included imperfections do not strongly impact the 
 waist and that the Super-Gaussian fit of the intensity is good enough to reproduce most of the basic features of the laser in the far field (see figure \ref{Near-far-profiles}).
Therefore, only the  introduction of the measured wave front and  imperfections in the intensity profile are responsible for the differences observed in the results.

\subsection{Numerical modeling}\label{numeric}

Particle-In-Cell is a powerful method that gives an accurate description of the plasma response to the laser and captures a wide range of physical phenomena \cite{Birdsall2004}. Nevertheless, precise and realistic results are obtained only with full 3D descriptions and high resolutions. Even though 2D simulations are used in the context of 2D
Cartesian slab or in the cylindrical geometry r-z to illuminate the physics, there is a qualitative and quantitative difference with the 3D simulations especially in the case of LWFA when studying
nonlinear regime. This is mainly because self-focus and self-modulation of the phase are not well described by a 2D simulation \cite{Davoine2008}. In the context of realistic simulations, this becomes particularly acute because of laser inhomogeneities and asymmetry.

The 3D cartesian PIC simulations with the different transverse laser profiles  are carried out using the code \Smilei \cite{Smilei2018} in a moving window of $6400 \times 640 \times 640$ cells of dimensions of $0.1~ c/ \omega_0$ in the longitudinal direction and $2.5 ~c / \omega_0$ in the transverse direction and 4 particles per cell. The simulations start with the center of the laser at $z= f -5 \times 10^{-4}~ \rm m$ so that the focal plane is situated at the end of the $500~ \mu m$ rising density ramp of the pre-ionized  plasma plateau of an electronic density $n_e= 8.6 \times 10^{17} $ cm$^{-3}$. The simulation results are compared up to $\rm 1.9 ~ mm$ of propagation corresponding to $ \sim 2$ Rayleigh length ($Z_r  \simeq 0.94~ \rm mm$). This distance is limited by the  end of the self-guiding. 

\section{Results}\label{results}

\begin{figure}[!h]
\centering
\includegraphics[width=\linewidth]{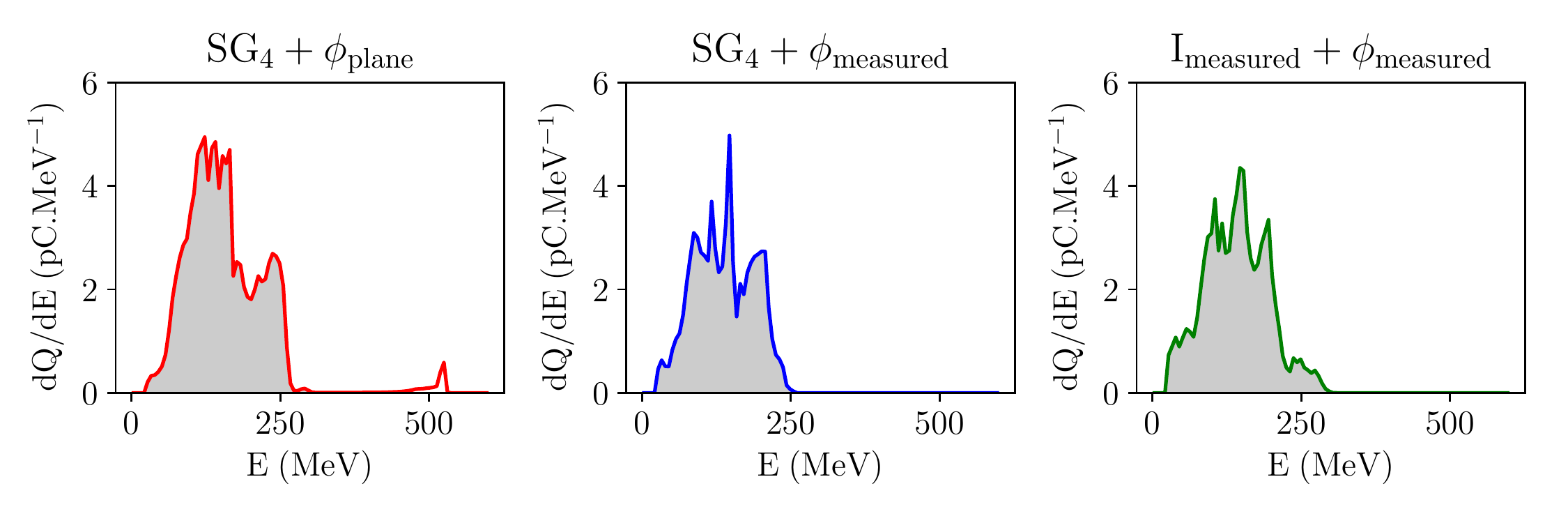}
\caption{Energy spectrum of the electron bunches, after $1.9~ mm$ of propagation. The shaded areas in grey  correspond to the selected parts  for which the bunch-related quantities are evaluated and compared.}
\label{compare-dist}
\end{figure}

The energy spectrum of the trapped electrons after a propagation distance of $\rm 1.9~ mm$ is shown in figure \ref{compare-dist}. The first remarkable difference between them is the presence of a low charge  high energy bunch of electrons only  in the case of $\mathrm{SG}_{4}+\phi_{\mathrm{flat}}$  which is originating from an early longitudinal injection \cite{CordeNatComm2013}.
In order to accurately compare the quantities related to the different resulting bunches, this first longitudinal injection is neglected  as it entirely vanishes in the two other simulations. In the following analysis related to the bunch quality, the quantities are evaluated only for the main high charge bunches designated by  the grey shaded  area in figure \ref{compare-dist}.

\subsection{Laser guiding}

\begin{figure}[!h]
\centering
\includegraphics[width=\linewidth]{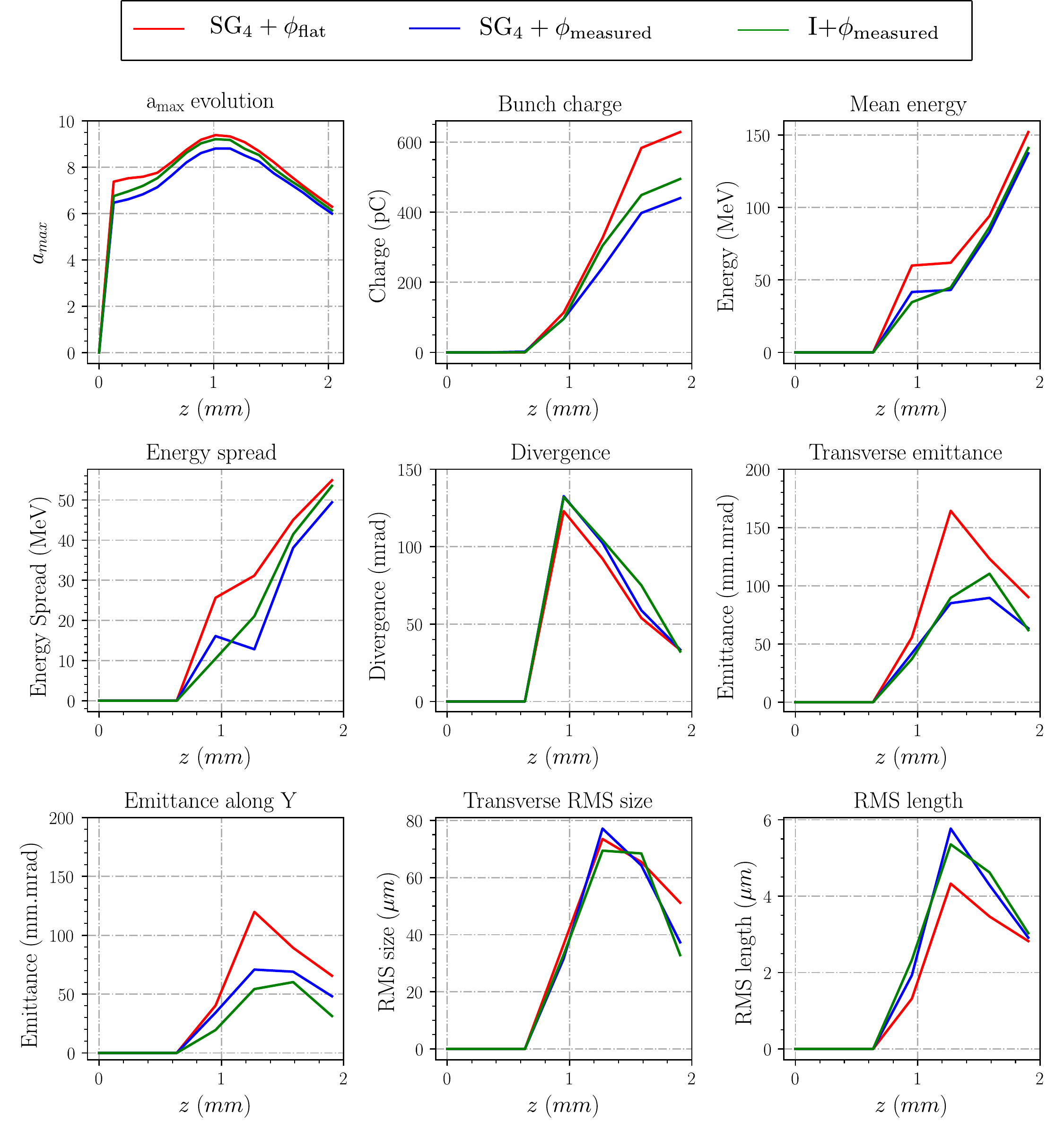}
\caption{The \textit{top left panel} shows the evolution of  the peak normalized laser field strength $a_{max}$. The other panels show the evolution of the bunch-related quantities  as a function of the propagation distance $z$.}
\label{Bunch-prop}
\end{figure}

The guiding of the different lasers is examined by comparing the evolution of  the peak normalized laser field strength $a_{max}$ as a function of the propagation distance (first panel of figure \ref{Bunch-prop}). In all cases, the maximal intensity first increases when the laser self-focuses at the entrance of the plasma, then decreases when the laser is defocused. The laser is self-guided over an acceleration length of $1.9~ mm$ and diffracts later on in these simulations. 

The self-focusing process begins approximately at the same time for the different profiles. Nevertheless, $a_0$ the initial value of $a_{max}$ is lower when the experimental features are introduced ($a_0 \simeq 7.5$ for $\mathrm{SG}_4 +\phi_{\mathrm{flat}}$ , $6.5$ for  $\mathrm{SG}_4+ \phi_{\mathrm{measured}}$  and $6.75$ for  I$ +\phi_{\mathrm{measured}}$).
The transverse laser profiles in the far field  presented in figure \ref{Near-far-profiles} being different, the laser intensity peak is also different in these three cases.

This difference is due to the diffraction of a portion of the laser energy in the wings, out of the central spot, when it propagated  from the near field to the far field. Thus, this fraction of energy does not contribute initially to the maximal intensity leading to a less important initial $a_{max}$.

With the same initial power in the near field  and a  similar spatial distribution of the laser in far field, the small difference in $a_0$ between $\mathrm{SG}_4+ \phi_{\mathrm{measured}}$ and I$+\phi_{\mathrm{measured}}$ is explained by a  compensation of the diffracted portion of the laser power in the wings introduced by the inhomogeneities in the intensity profile. In fact, in the case of I$+\phi_{\mathrm{measured}}$ the correctly focused area from the near field carries on average more intensity  compared to the case of  $\mathrm{SG}_4+ \phi_{\mathrm{measured}}$.

Due to the introduction of the realistic wavefront and intensity profile, the  laser self-focusing evolution is modified. Despite the difference in the initial value of $a_{max}$, the laser self-focusing is quicker  when the measured laser wavefront is introduced in either  $\mathrm{SG}_4+ \phi_{\mathrm{measured}}$ or I$+\phi_{\mathrm{measured}}$ (note the difference in the slope). However, the slope becomes steeper in the case of I$+\phi_{\mathrm{measured}}$ where both the wavefront and the intensity are combined. 
In fact, the initial gap in the intensity between 
$\mathrm{SG}_4 +\phi_{\mathrm{flat}}$ and the realistic profiles is gradually reduced during the self-focusing, suggesting that a part of the energy contained in the outer part is actually self-focused and guided by the plasma. Therefore, it eventually contributes to the physical processes involved in LWFA like the bubble formation and electron acceleration.
This mechanism explains why differences in some of the beam quantities remain relatively small at the end of the propagation.
The maximal intensity with realistic features reaches a comparable, yet slightly inferior, value to the one found with a flat wavefront. The effect of the portion of the laser pulse energy situated out of the focal spot, referred to as \textit{halo effect} has been investigated experimentally and numerically in \cite{Nakanii2016}. The  halo of the pulse can  be self-focused and  thus can also contribute to the self-injection process, given that it has enough power itself. 

The gap in the initial $a_{max}$  between $\mathrm{SG}_4+ \phi_{\mathrm{measured}}$ and I$+\phi_{\mathrm{measured}}$  is conserved during the focusing phase of the laser. However, it is slightly reduced during the defocusing phase.   This  suggests that  the wavefront effect is dominant compared to imperfections related  to the intensity profile. This is also reflected in the beam quantities as it is discussed in the next section. 

\subsection{Beam properties}

As mentioned earlier, the bunch-related quantities presented in \ref{Bunch-prop} are evaluated on the main high-charge electron bunches that are shaded in grey in figure \ref{compare-dist}.

Because of the lower intensity peak in the realistic profiles during their propagation in the plasma, the effective energy transferred from the laser into the plasma is less important resulting in the generation of a wakefield with a lower amplitude. Even though the energy in the wings of  the laser spot is  partially self-focused, the rest of it is  actually wasted and  not coupled into the plasma wave \cite{Genoud2013}. Therefore, the quantity of the injected charge is less important when the imperfections of the experimental laser are added. In fact, the bunch charge in the case of $\mathrm{I}+\phi_{\mathrm flat}$ is $\sim 35 \%$ less than that in $\mathrm{SG}_4+\phi_{\mathrm{flat}}$ case. Accordingly, the mean energy of the bunch is also less important which shows that the additional charge does not trigger significant beam loading in this case.

\begin{figure}
     \centering
     \begin{subfigure}[b]{0.41\textwidth}
         \centering
         \includegraphics[width=\textwidth]{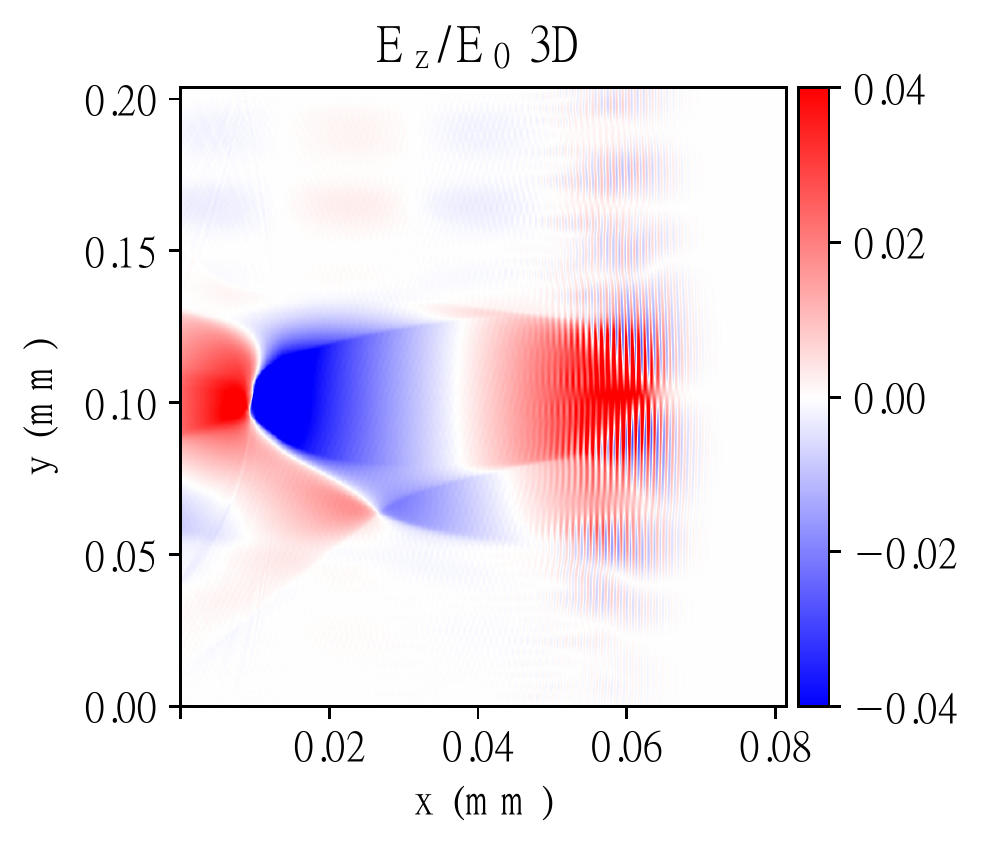}
         \caption{Normalized longitudinal electric field}
         \label{fig:realisticEz}
     \end{subfigure}
     \begin{subfigure}[b]{0.4\textwidth}
         \centering
         \includegraphics[width=\textwidth]{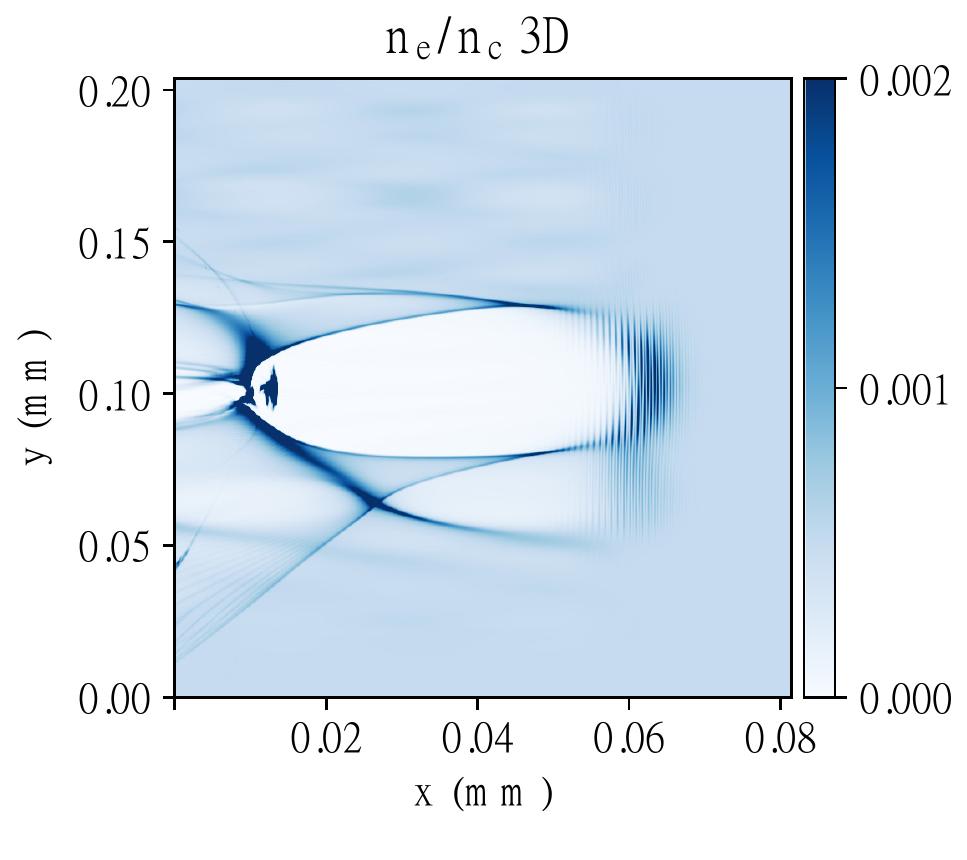}
         \caption{normalized electronic density}
         \label{fig:realisticRho}
     \end{subfigure}
        \caption{Colormaps ofthe normalized longitudinal electric field and the normalized electronic density colormaps in 3D simulation of the case of a realistic laser pulse with measured intensity and phase after 1.9 mm of propagation.}
        \label{fig:realisticlaser}
\end{figure}

Since the simulations are carried in the blow-out regime, electrons are almost completely expelled from the region behind the laser pulse around its propagation axis. As we can see in figure \ref{fig:realisticlaser}, a thin sheath of electrons creates a separation between  the ion cavity and  the surrounding plasma behind the pulse and it bends the laser-pulse wavefronts outwards for $r  \gtrsim w_0  \sim R_b$, and inwards for $r \lesssim w_0  \sim R_b$, where $R_b$ is the maximal radius of the bubble \cite{Vieira2012}. Hence, the blowout region acts as a spatial filter of the laser pulse: only the main pulse is focused and plays a role in the wake generation mechanism,  while the rest of the pulse is lost via diffraction and pump depletion and does not contribute to electron acceleration. 

 Besides, by looking at the different quantities related to the bunch properties in figure \ref{Bunch-prop} around $1.6~ mm$ of acceleration, one can see that the laser imperfections lead to some differences in the quality  of the bunch. The inhomogeneities in the laser pulse lead to a distorted wakefield whose focusing and defocusing properties may affect the electron distribution. As a consequence, the corresponding  distribution of accelerated electrons is less collimated   than that in the case of $\mathrm{SG}_4+\phi_{\mathrm{flat}}$. Despite the lower injected charge, the realistic profiles lead to a slightly higher value of divergence.

Counter-intuitively, the laser imperfections lead to an enhanced transverse beam emittance, particularly in the $y-$direction where the profiles are more distorted in the far field. This has been also observed in the simulations conducted in \cite{Cummings2011} where the wings are referred to as coma aberration. This enhancement  can be explained by the difference in the quantity of injected electrons. 
The aberrations introduced by the experimental profiles also drive a longer bunch yet  smaller  transversely.

The differences in the quantities indicating the quality of the bunch  are considerably reduced after approximately $1.9~ \rm mm$ of acceleration, especially between the $\mathrm{SG}_4+ \phi_{\mathrm{measured}}$ and I$+\phi_{\mathrm{measured}}$. This puts forward the sensitivity of the  electron beam distribution to the laser wavefront rather than the intensity profile itself  \cite{Beaurepaire2015}.  
In fact, by examining closely the transverse laser profiles in the far field of $\mathrm{SG}_4+ \phi_{\mathrm{measured}}$ and I$+\phi_{\mathrm{measured}}$, there is an important resemblance between the far field  laser profiles in the two cases. Therefore, most of the  spatial inhomogeneities  are rather dictated  by the laser wavefront distortions.

\begin{figure}
\begin{subfigure}{\textwidth}
  \centering
  \caption{}
  \includegraphics[width=\linewidth]{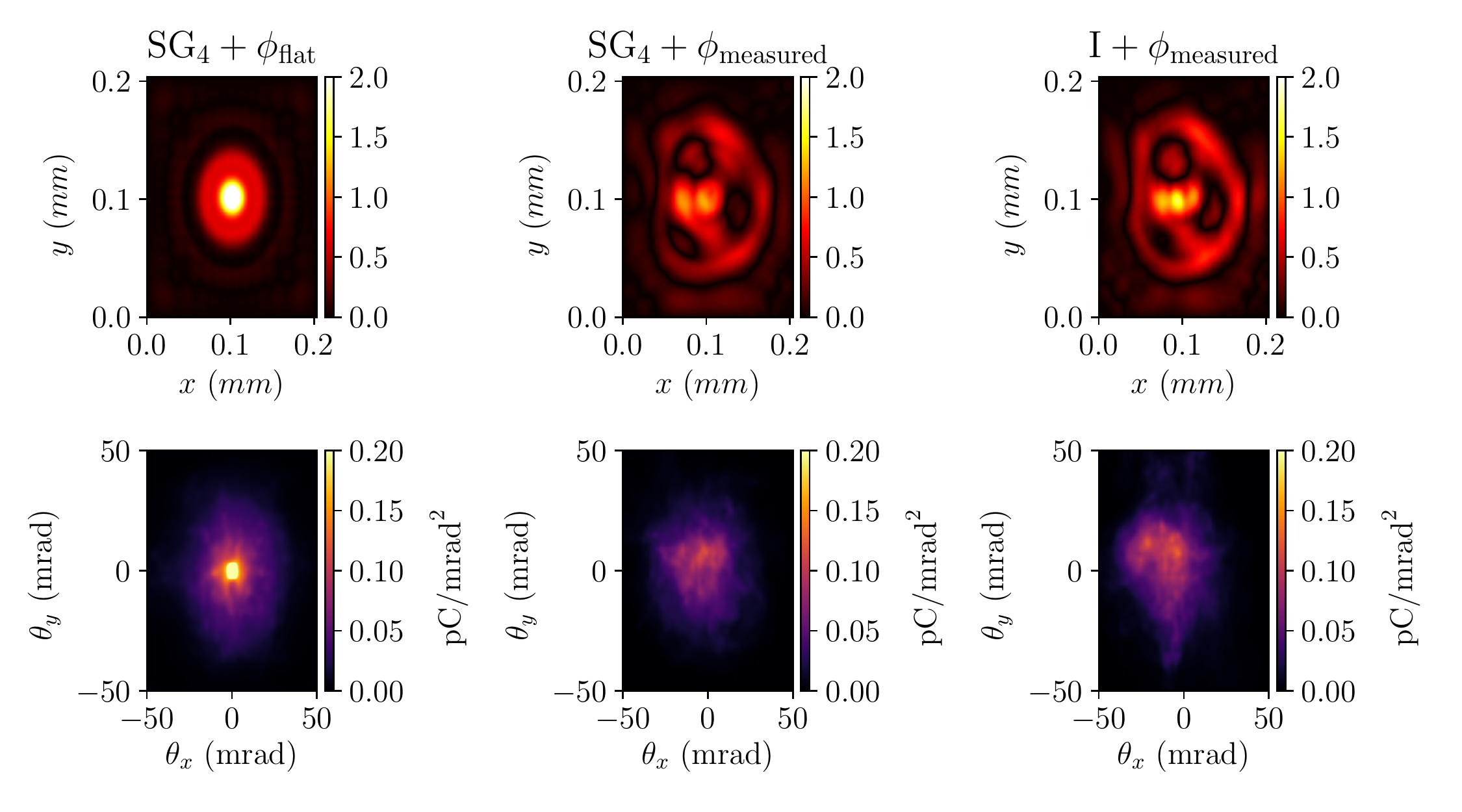}  
\label{compare-bunch1}
\end{subfigure}
\newline
\begin{subfigure}{\textwidth}
  \centering
  \caption{}
  \includegraphics[width=\linewidth]{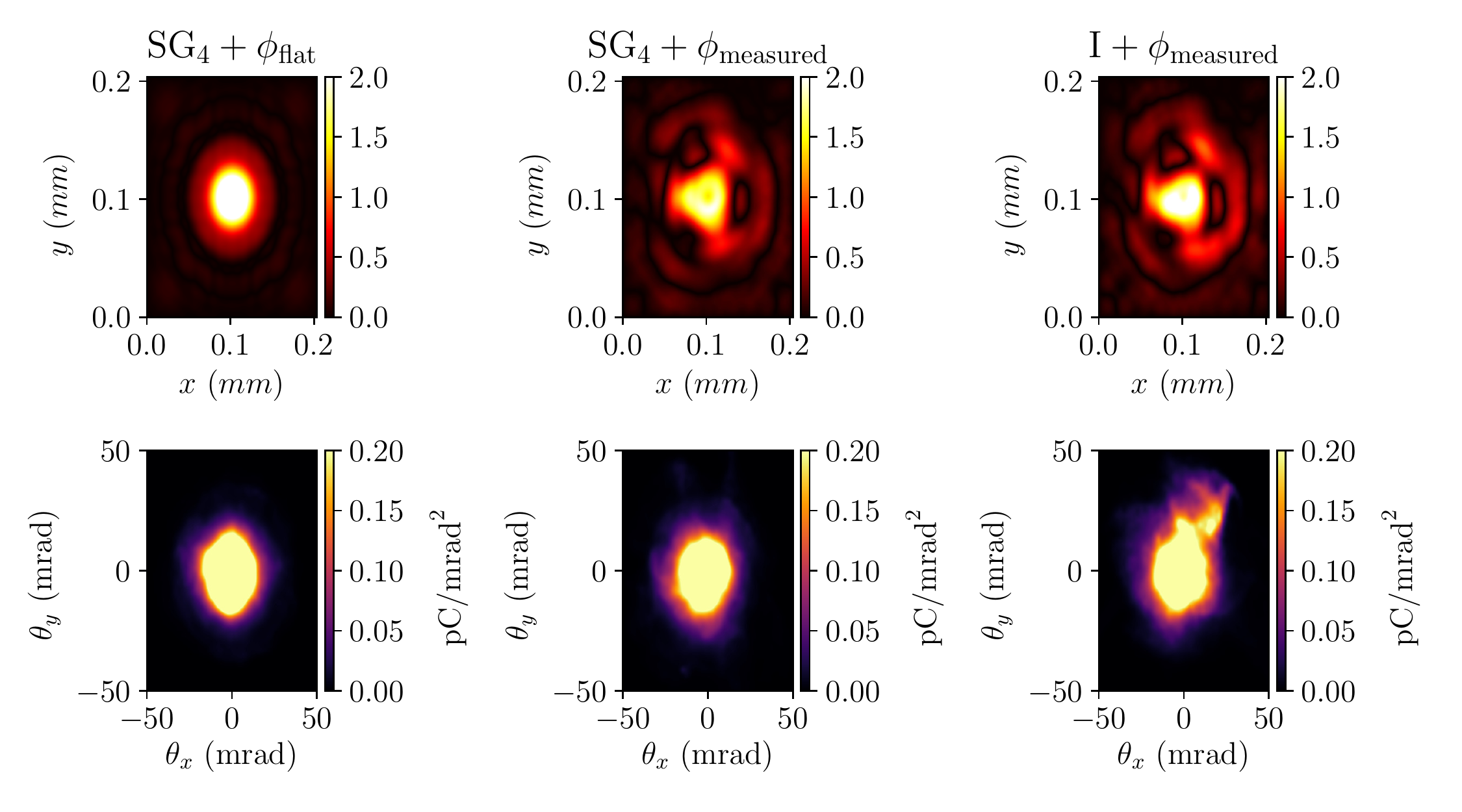}  
\label{compare-bunch2}
\end{subfigure}
 \caption{From left to right: simulations results for the $\mathrm{SG}_4 +\phi_{\mathrm{flat}}$, $\mathrm{SG}_4+ \phi_{\mathrm{measured}}$ and  I$+\phi_{\mathrm{measured}}$. The  \textit{top panels} are the   cross-sections of the  normalized transverse fields $E_{\perp}/ E_0$ where $E_0= m_e c \omega_0 /e$ and the  \textit{lower panels} are the distribution of the electron bunches in $(\theta_x, \theta_y)$ in (a) after $1.2~ mm$ and in (b) after $1.9~mm$ of propagation.}
\end{figure}

It also demonstrates the role of the plasma to filter the spatial asymmetries  of the transverse wakefield by homogenizing it during laser propagation.  Indeed, the halo of the pulse once self-focused gives rise to its own wakefield. Because of the difference in the size between the halo and the main laser pulse,  the  angle of the generated wakefield  is different and the interaction between the different wakefields induces some instability  at the beginning.
However, since the energy contained in the wings is way lower than that of the central part, it is quickly depleted in the plasma leading to a smoother propagation afterward when the fraction of this energy is dissipated in the plasma. Therefore, the broad distribution of the electron beam becomes more collimated  and its divergence is decreased under the effect of the strong self-focusing field inside the bubble. This is illustrated in figures \ref{compare-bunch1} and \ref{compare-bunch2} which show  the normalized  transverse laser field $E_{\perp}/E_0$  with $E_{\perp}=\sqrt{E_x^2+ E_y^2}$  and $E_0= m_e c \omega_0/ e$ and the transverse electron distributions in $(\theta_x, \theta_y)$ for respectively $1.2~mm $ and $1.9~ mm$ of propagation.

\ref{compare-bunch2} shows that the aberrations in the transverse wakefield are self-corrected by the plasma response, leading to a stable propagation and  little deviation in the electron distributions.

\section{Discussion and conclusion}\label{discussion}

In this  study, the effects of  realistic laser profiles  have been numerically  investigated  where both the experimental intensity distribution and wavefront are used as input in the three-dimensional PIC simulations. The results are compared with a  super-Gaussian profile of order 4 along with a perfectly flat wavefront at the focal plane.

 Keeping the total laser energy constant, the introduction of realistic features results in a significant drop in the injected charge and a decrease of the mean energy of the beam. This clearly emphasizes the limitation of using standard laser profiles in simulations and put forward in particular the importance of the wavefront which can be deleterious to the results. It has been shown that  the wavefront distortions  in the near field play an important role in shaping the asymmetries and inhomogeneities present in the laser's far field\cite{Leemans2014}.  These imperfections in the laser pulse result in a complex wakefield pattern that influences the electrons injection process and their quality in return.  
 
 This suggests  the possibility of boosting the electron bunch quality by properly  tailoring the laser wavefront, which can be achieved by  using a deformable mirror as proposed in \cite{He2013,HeZ2015}. 
Presently, the main solution to increase  the electron charge or energy  relies on rising the laser power. However, this analysis emphasizes the benefit of improving the quality of the laser in the far field \cite{Genoud2013, Mangles2012}. 
 By controlling the wavefront, the pulse energy contained within the central part of the focal spot can be increased while keeping the total energy constant resulting in  a better coupling with the plasma and a more efficient energy conversion to the accelerated electrons.  

It also has been demonstrated that the plasma plays an important role to focus and guide a portion of the laser energy located in the wings and reduces the inhomogeneities doing so. In \cite{Ferri2016}, this role is highlighted by examining the  position of the focal plane with respect to the plasma. A better guiding is found by moving the focal plane further inside the plasma. This can be explained by the fact that the smoothing action of the plasma is enhanced by increasing the laser-plasma interaction length before the full focalization. This eventually leads to a more homogeneous focal spot.

\ack
This work was granted access to the HPC resources of Jean Zay under the Grand Challenge project allocation 100953 made by GENCI. 
The authors are also thankful  to the experimenters who are working on the Apollon laser for providing  the experimental data and the laser measurements  used for this numerical study, in particular, Dimitris Papadopoulos and Mélanie Chabanis.

\section*{References}

\bibliography{Bibliography}

\providecommand{\noopsort}[1]{}\providecommand{\singleletter}[1]{#1}%
\providecommand{\newblock}{}
\begin{thebibliography}{10}
\expandafter\ifx\csname url\endcsname\relax
  \def\url#1{{\tt #1}}\fi
\expandafter\ifx\csname urlprefix\endcsname\relax\def\urlprefix{URL }\fi
\providecommand{\eprint}[2][]{\url{#2}}

\bibitem{Sun87}
Sun G, Ott E, Lee Y~C and Guzdar P 1987 {\em The Physics of Fluids\/} {\bf 30}
  526--532

\bibitem{Benedetti2013}
Benedetti C, Schroeder C~B, Esarey E, Rossi F and Leemans W~P 2013 {\em Physics
  of Plasmas\/} {\bf 20} 103108

\bibitem{Malka2013proc}
Malka V 2013 {Review of Laser Wakefield Accelerators} {\em {4th International
  Particle Accelerator Conference}\/}

\bibitem{Kaluza2010}
Kaluza M~C, Mangles S~P~D, Thomas A~G~R, Najmudin Z, Dangor A~E, Murphy C~D,
  Collier J~L, Divall E~J, Foster P~S, Hooker C~J, Langley A~J, Smith J and
  Krushelnick K 2010 {\em Phys. Rev. Lett.\/} {\bf 105}(9) 095003

\bibitem{Glinec2008}
Glinec Y, Faure J, Lifschitz A, Vieira J~M, Fonseca R~A, Silva L~O and Malka V
  2008 {\em Europhysics Letters\/} {\bf 81} 64001

\bibitem{Ferri2016}
Ferri J, Davoine X, Fourmaux S, Kieffer J~C, Corde S, Ta~Phuoc K and Lifschitz
  A 2016 {\em Scientific Reports\/} {\bf 6} 27846

\bibitem{Vieira2012}
Vieira J, Martins S~F, Fi{\'{u}}za F, Huang C~K, Mori W~B, Mangles S~P~D, Kneip
  S, Nagel S, Najmudin Z and Silva L~O 2012 {\em Plasma Physics and Controlled
  Fusion\/} {\bf 54} 055010

\bibitem{Michel2006}
Michel P, Esarey E, Schroeder C~B, Shadwick B~A and Leemans W~P 2006 {\em
  Physics of Plasmas\/} {\bf 13} 113112

\bibitem{Mangles2009}
Mangles S~P~D, Genoud G, Kneip S, Burza M, Cassou K, Cros B, Dover N~P,
  Kamperidis C, Najmudin Z, Persson A, Schreiber J, Wojda F and Wahlström C~G
  2009 {\em Applied Physics Letters\/} {\bf 95} 181106

\bibitem{Genoud2013}
Genoud G, Bloom M~S, Vieira J, Burza M, Najmudin Z, Persson A, Silva L~O,
  Svensson K, Wahlström C~G and Mangles S~P~D 2013 {\em Physics of Plasmas\/}
  {\bf 20} 064501

\bibitem{Nakanii2016}
Nakanii N, Hosokai T, Iwasa K, Pathak N~C, Masuda S, Zhidkov A~G, Nakahara H,
  Takeguchi N, Mizuta Y, Otsuka T~P, Sueda K, Nakamura H and Kodama R 2016 {\em
  Europhysics Letters\/} {\bf 113} 34002

\bibitem{Maslarova2019}
Maslarova D, Krus M, Horny V and Psikal J 2019 {\em Plasma Physics and
  Controlled Fusion\/} {\bf 62} 024005

\bibitem{Kirchen2021}
Kirchen M, Jalas S, Messner P, Winkler P, Eichner T, H\"ubner L, H\"ulsenbusch
  T, Jeppe L, Parikh T, Schnepp M and Maier A~R 2021 {\em Phys. Rev. Lett.\/}
  {\bf 126}(17) 174801

\bibitem{Gori1994}
Gori F 1994 {\em Optics Communications\/} {\bf 107} 335--341 ISSN 0030-4018

\bibitem{Cummings2011}
Cummings P and Thomas A~G~R 2011 {\em Physics of Plasmas\/} {\bf 18} 053110

\bibitem{Birdsall2004}
Birdsall C and Langdon A 2004 {\em Plasma Physics via Computer Simulation\/}
  Series in Plasma Physics (Taylor \& Francis) ISBN 9780750310253

\bibitem{Davoine2008}
Davoine X, Lefebvre E, Faure J, Rechatin C, Lifschitz A and Malka V 2008 {\em
  Physics of Plasmas\/} {\bf 15} 113102

\bibitem{Smilei2018}
Derouillat J, Beck A, Pérez F, Vinci T, Chiaramello M, Grassi A, Flé M,
  Bouchard G, Plotnikov I, Aunai N, Dargent J, Riconda C and Grech M 2018 {\em
  Computer Physics Communications\/} {\bf 222} 351 -- 373

\bibitem{CordeNatComm2013}
Corde S, Thaury C, Lifschitz A, Lambert G, Ta~Phuoc K, Davoine X, Lehe R,
  Douillet D, Rousse A and Malka V 2013 {\em Nat Commun\/} {\bf 4}

\bibitem{Beaurepaire2015}
Beaurepaire B, Vernier A, Bocoum M, B\"ohle F, Jullien A, Rousseau J~P, Lefrou
  T, Douillet D, Iaquaniello G, Lopez-Martens R, Lifschitz A and Faure J 2015
  {\em Phys. Rev. X\/} {\bf 5}(3) 031012

\bibitem{Leemans2014}
Leemans W~P, Gonsalves A~J, Mao H~S, Nakamura K, Benedetti C, Schroeder C~B,
  T\'oth C, Daniels J, Mittelberger D~E, Bulanov S~S, Vay J~L, Geddes C~G~R and
  Esarey E 2014 {\em Phys. Rev. Lett.\/} {\bf 113}(24) 245002

\bibitem{He2013}
He Z~H, Hou B, Nees J~A, Easter J~H, Faure J, Krushelnick K and Thomas A~G~R
  2013 {\em New Journal of Physics\/} {\bf 15} 053016

\bibitem{HeZ2015}
He Z~H, Hou B, Lebailly V, Nees J~A, Krushelnick K and Thomas A~G~R 2015 {\em
  Nature Communications\/} {\bf 6} 7156

\bibitem{Mangles2012}
Mangles S~P~D, Genoud G, Bloom M~S, Burza M, Najmudin Z, Persson A, Svensson K,
  Thomas A~G~R and Wahlstr\"om C~G 2012 {\em Phys. Rev. ST Accel. Beams\/} {\bf
  15}(1) 011302

\end{thebibliography}
\bibliographystyle{iopart-num}

\end{document}